\documentclass[10pt, conference, compsocconf]{IEEEtran}

\usepackage{algorithm}
\usepackage{algpseudocode}
\usepackage{multirow}
\usepackage{graphicx}
\usepackage[bookmarks=false]{hyperref}
\usepackage{cite}
\usepackage{epsfig}
\usepackage{amsmath, amssymb}
\usepackage{wrapfig}
\usepackage{epstopdf}
\usepackage{caption}
\usepackage{color}
\usepackage{colortbl}
\usepackage{subfigure}
\usepackage{indentfirst}
\usepackage{url}
\usepackage{mathptmx}
\usepackage{subfigure}
\graphicspath{{figures/}}
\usepackage[autostyle]{csquotes}
\usepackage{balance}
\usepackage{kotex}

\begin{document}

\title{CloudSafe: A Tool for an Automated Security Analysis for Cloud Computing}

\author{\IEEEauthorblockN{Seongmo An\IEEEauthorrefmark{1}, Taehoon Eom\IEEEauthorrefmark{1}, Jong Sou Park\IEEEauthorrefmark{1}, Jin B. Hong\IEEEauthorrefmark{2},\\ Armstrong Nhlabatsi\IEEEauthorrefmark{3}, Noora Fetais\IEEEauthorrefmark{3}, Khaled M. Khan\IEEEauthorrefmark{3}, Dong Seong Kim\IEEEauthorrefmark{4}}
\IEEEauthorblockA{\IEEEauthorrefmark{1}Dept. of Computer Engineering, Korea Aerospace University, South Korea\\
\IEEEauthorrefmark{2}Department of Computer Science and Software Engineering, University of Western Australia, Australia\\
\IEEEauthorrefmark{3}KINDI Center for Computing Research, Qatar University, Qatar\\
\IEEEauthorrefmark{4}School of Information Technology and Electrical Engineering,
University of Queensland, Australia\\
seongmoa@kau.ac.kr, eomth86@kau.ac.kr. jspark@kau.ac.kr, jin.hong@uwa.edu.au\\
Armstrong.Nhlabatsi@qu.ed.qa, N.Fetais@qu.edu.qa, K.Khan@qu.edu.qa, dan.kim@uq.edu.au}
}

\maketitle

\begin{abstract}
Cloud computing has been adopted widely, providing on-demand computing resources to improve perfornance and reduce the operational costs. However, these new functionalities also bring new ways to exploit the cloud computing environment. To assess the security of the cloud, graphical security models can be used, such as Attack Graphs and Attack Trees. However, existing models do not consider all types of threats, and also automating the security assessment functions are difficult. In this paper, we propose a new security assessment tool for the cloud named \textit{CloudSafe}, an automated security assessment for the cloud. The CloudSafe tool collates various tools and frameworks to automate the security assessment process. To demonstrate the applicability of the CloudSafe, we conducted security assessment in Amazon AWS, where our experimental results showed that we can effectively gather security information of the cloud and carry out security assessment to produce security reports. Users and cloud service providers can use the security report generated by the CloudSafe to understand the security posture of the cloud being used/provided.
\end{abstract}


\section{Introduction}
\label{intro}
Cloud computing provides many beneficial factors over the traditional networks, which enhance the productivity and performance of enterprises and individuals \cite{mell2011cloud, bani2013meridian}. But at the same time, it faces many security challenges and threats, affecting the decisions of using the cloud computing services significantly \cite{hong2019systematic}. That is, potential cloud users need to consider various security implications before migrating their data and operations to the cloud computing environment \cite{Gonzales2017}. Although there are various security mechanisms implemented for the cloud, their effectiveness must be evaluated to fully understand the security posture of the cloud. One of the widely used techniques is to develop a security assessment framework using graphical security models \cite{kordy2013dag, hong2017survey}. These models provide the framework to collect security data and evaluate various attack scenarios of the network, as well as the capabilities to incorporate countermeasure selections. However, automating the functionalities of those models in the cloud can be a challenge, as the privilege boundaries are more complex than the traditional networks.

Currently, much of the security analysis for the cloud is done manually by a security expert. In this process, a lot of time is consumed and it can also introduce human errors. Hence, automation is needed to reduce the cost and time, as well as reducing human errors. There are many tools available for assessing the security of networks \cite{hong2017survey}. For example, NAVIGATOR \cite{chu2010visualizing}, MulVal \cite{ou2005mulval}, and NICE \cite{chung2013nice} all have functions to automatically assess the security of networks. However, these tools require specific inputs, which are security details of the network. Security information gathering tools, such as NESSUS \cite{nessus} and National Vulnerability Database (NVD) \cite{nvd}, are easily deployed in traditional networks as typically the whole network is under the system administrator's control. However, the cloud separates the privilege between different stakeholders (e.g., clients and service providers), limiting the access to security information that existing information gathering tools can access to. Hence, additional measures are needed to ensure that all these information can be collected automatically to evaluate the security of the cloud.

To address the aforementioned problems, we propose an automated cloud-based security analysis framework to evaluate the security of the cloud. We present additional techniques to collect security information from the cloud, which is then stored in a database. As the security assessment framework in the CloudSafe, we implement a scalable graphical security model named Hierarchical Attack Representation Model (HARM) \cite{hong2016towards}. We further modify the functionalities of the HARM such that it will integrate with the security data gathering interfaces we implemented for the CloudSafe. Finally, we carry out experimental analysis in Amazon AWS to validate the applicability and practical use of the CloudSafe. The contributions of the paper are as follows.

\begin{itemize}
\item To develop security information gathering interfaces for the cloud given privilege separation boundaries,
\item To implement the CloudSafe by integrating the functionalities of security information gathering interfaces and the HARM,
\item To conduct experiments in the Amazon AWS to evaluate the applicability and practical use of the CloudSafe.
\end{itemize}

The rest of the paper is structured as follows. Section ~\ref{back} describes the related work on security assessment for the cloud. Section ~\ref{proposal} presents the CloudSafe framework with details on its architecture, configurations and workflow. Section ~\ref{result} shows the results of using the CloudSafe in the Amazon AWS. Finally, we  conclude the paper in Section ~\ref{conclusion}.

\section{Related Work}
\label{back}
The CloudSafe is a security assessment tool for the cloud, which utilizes cloud resources to perform its tasks. In this section, we review existing methods for evaluating the security of the cloud, and also the techniques of utilizing cloud resources for the security asssessments. 

\subsection{Cloud-Based Security Assessment}
The cloud provides scalable and on-demand computing resources, which can be utilized for the security assessments. Martinez \emph{et al}. ~\cite{Martinez2010} proposed an architecture for malware detection based on the concept of web services and the malware intrusion ontology. Using multiple engines that implement heterogeneous analysis strategies, it has detected malicious content or behavior in unknown files. Ouellette \emph{et al}.~\cite{Ouellette2013} proposed a cloud-based malware detection learning algorithm. The algorithm detects malware by looking for properties that cannot change in malware such as known patterns of malware behavior. Shin and Gu ~\cite{shin2012} proposed ClOUDWATCHER - a network monitoring framework using OpenFlow in a dynamic cloud network. Chen \emph{et al}.~\cite{chen2013} proposed collecting and storing cloud operational reports, security events, and traffic data and using parallel forensic analysis. Khune and Thangakumar ~\cite{khune2012} are cloud-based developers that created the virtualized and synchronized replicas of real smart phones in a cloud environment. Multiple detection engines are run in parallel to detect attacks, smart phone intrusion. Mahmood \emph{et al}. ~\cite{Mahmood2012} proposed a framework for performing security tests on smart phones using fuzzy testing in a cloud environment. As demonstrated by those approaches, the cloud can provide substantial benefit for the security assessments. However, almost all graphical security models do not utilize the cloud computing resources \cite{hong2017survey}. To improve the performance and management of security assessment tasks, we develop a framework to utilize the cloud computing resource for graphical security models in the proposed SafeCloud tool.

\subsection{Security Assessment of the Cloud} 
Bleikertz \emph{et al}. ~\cite{Bleikertz2010} proposed a query and policy language that could be used to specify desired and unwanted configurations in the network. When an attack query is entered, it indicates whether an attack route exists based on the attack graph. Noel \emph{et al}. ~\cite{noel2016cygraph} proposed a scalable modeling framework to create a predictive model for possible multi-step attack paths by combining vulnerability, server configuration, policy rules, and security events. The generated attack path is stored in the Neo4j-based database and provides the result of combining with the input query. Rizvi \emph{et al}. \cite{rizvi2018security} proposed a framework for evaluating the security of the cloud. They presented security evaluation rules to assess the security using the developed security metrics (based on linear and non-linear equations and fuzzy logic systems). However, they only covered specific aspects of security, such as interoperability, co-location, transparency, malicious insider and portability. Manzoor \emph{et al}. \cite{manzoor2018threat} proposed a new security assessment model for the cloud using Petri Nets. They profile the operational behavior of the services in the cloud operations, which are then used to evaluate the security of the cloud operations in different layers. However, using Petri Nets can have a scalability problem, especially for the cloud where configurations can dynamically change within a short period of time. There are many other graphical security models that could be used to assess the security of the cloud \cite{hong2017survey, kordy2013dag}. However, we must first specify how the security assessment could be carried out in the cloud environment ensuring the data collection, processing and evaluation, which has not been specified.

\section{CloudSafe}
\label{proposal}

\begin{figure*}
\center{\epsfig{file=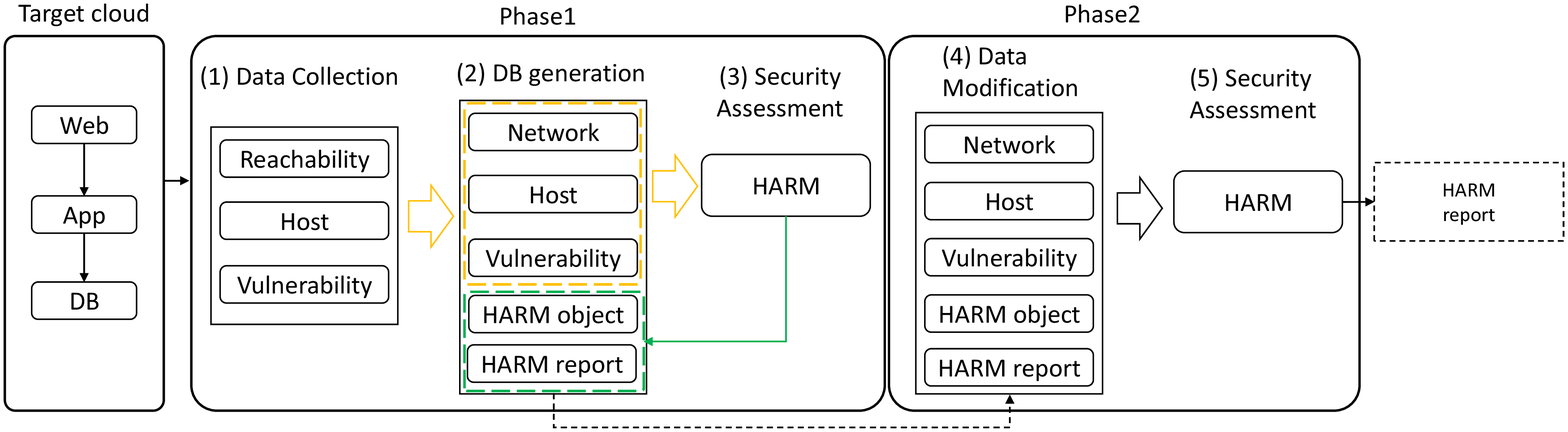, width=0.9\textwidth,clip=}}
\caption{Framework Architecture}
\label{fig:Framework} 
\end{figure*}

The details of the CloudSafe tool is presented in this section. CloudSafe follows the SaaS (Software-as-a-Service) framework that can perform cloud security analysis (i.e. a cloud service for cloud security). Figure ~\ref{fig:Framework} shows the overall process of the CloudSafe framework. The proposed framework is implemented in the Amazon AWS (AWS for short), and it consists of two phases. Phase 1 collects information about the target cloud and stores the data used for the security analysis in a database. Then, the security is evaluated using the HARM (Hierarchical Attack Representation Model). Phase 2 generates and stores a new HARM model by modifying the security information collected in Phase 1. In this way it can be seen if the security posture of the cloud changes without changing the actual cloud configurations.

\subsection{Phase1}
\begin{figure}
\center{\epsfig{file=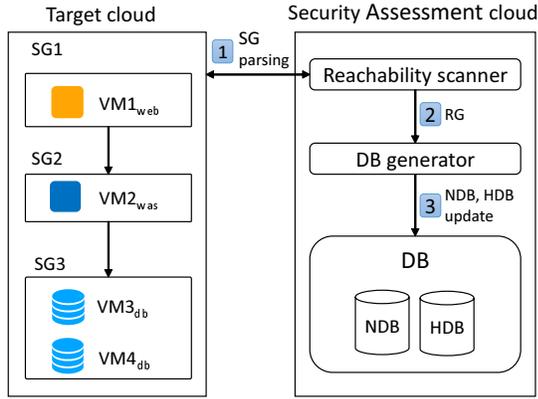, width=0.4\textwidth,clip=}}
\caption{Reachability scanning in Phase 1}
\label{fig:2} 
\end{figure}

Reachability of the cloud components in the security analysis is essential information (e.g., the reachability of different virtual machines (VMs)). There are various tools for this purpose, but it is difficult to apply  them in the cloud, because the assessment tool may not have enough privileges to access such information. To solve this problem, we obtained Reachability using Security Group (SG), which is the basic security method of AWS. The SG controls access using IP and Port as packet filtering. The SG information is used to generate the Reachability Graph (RG) by considering only the allowed rules for inbound traffics. Figure~\ref{fig:2} shows the process of acquiring the reachability information from the target cloud (i.e., the cloud environment to conduct security assessment) and storing it in the Network Database (NDB) and Host Database (HDB) by parsing the SG to understand the inter-VM reliability. Then, an RG is generated and stored in the NDB, and basic information of the host is generated in the HDB. Algorithm \ref{algorithm1} is used to populate the RG from SG, which uses information. The algorithm iteratively goes over the set of security rules to examine the reachability specified in the SG, and continuously adding the new set into the RG.

\begin{figure}
\center{\epsfig{file=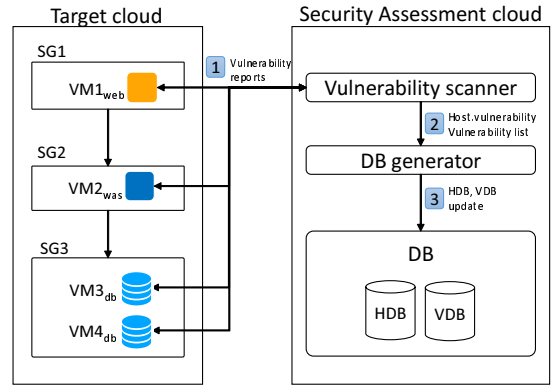, width=0.4\textwidth,clip=}}
\caption{Vulnerability scanning in Phase 1}
\label{fig:3} 
\end{figure}

Through the process shown in Figure~\ref {fig:2}, information about the reachability and VMs included in the cloud is collected and stored in the databases. However, the details of each VM are unknown. The details of the HDB are supplemented in the vulnerability scanning process. We use vulnerability analysis tools to find vulnerabilities in each VM and store them. This process is shown in Fig. ~\ref{fig:3}. The vulnerability analysis tool is used to scan all the VMs constituting the target cloud, and the information is stored in the HDB. Vulnerability scanning results include open ports, services provided, and vulnerabilities. Algorithm \ref{algorithm2} shows the process of collecting the vulnerability report. The module goes over every host (i.e., VMs) and conduct vulnerability scanning. Then, the corresponding report is stored into the vulnerability database (VDB). We use open port information and vulnerabilities of VMs. If there is no vulnerabilities in the VDB, the vulnerability information of the NVD (National Vulnerability Database \cite{nvd}) is retrieved and stored in the VDB.

\begin{algorithm}
\begin{algorithmic}
\Procedure{Generating~RG~from~SG}{$SG, NDB, HDB$}
\State{\texttt{Init $RG$}}
\For{\texttt{All~Security~Rule~in~SG}}
\If{\texttt{$RG \not\ni SG.host$}}
\State{\texttt{Add $SG.host~to~RG$}}
\State{\texttt{Insert $SG.host~to~RG$}}
\EndIf
\State{\texttt{Add $SG.reachability~to~RG$}}
\EndFor
\State{\texttt{Insert $RG~to~NDB$}}
\EndProcedure
\end{algorithmic}
\caption{Security Groups to Reachability Graph parser}
\label{algorithm1}
\end{algorithm}

\begin{algorithm}
\begin{algorithmic}
\Procedure{Analyse Vul Report}{$Report, HostID, HDB, VDB$}
\State{\texttt{Init $Host$}}
\State{\texttt{Insert $Time, OS, Ports to HDB(HostID)$}}
\For{\texttt{all vulnerabilities in report}}
\State{\texttt{$V \gets vulnerability$}}
\State{\texttt{Add $V~to~Host$}}
\If{\texttt{$VDB \not\ni V$}}
\State{\texttt{Insert $V~to~VDB$}}
\EndIf
\EndFor
\State{Insert $Host~to~HDB$}
\EndProcedure
\end{algorithmic}
\caption{Vulnerability scan report parser}
\label{algorithm2}
\end{algorithm}

\begin{figure}
\center{\epsfig{file=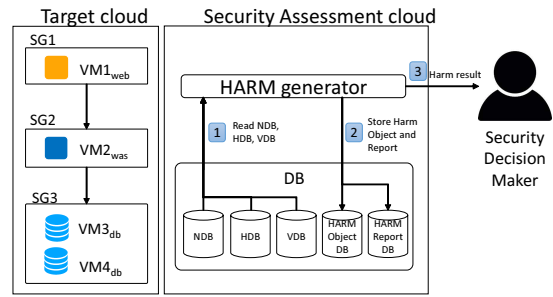, width=0.4\textwidth,clip=}}
\caption{Security assessment in Phase 1}
\label{fig:4} 
\end{figure}

With collected reachability and vulnerability data, we can generate the HARM as shown in Figure~\ref{fig:4}. The generated HARM is stored in the HARM object DB, which is then used to compare the changes in security of the cloud in phase 2. But this information could also be used later to carry out other off-line security assessments. Finally, the security analysis result from the HARM is provided to the user. 

\begin{algorithm}
\begin{algorithmic}
\Procedure{Generating HARM model from DB}{$NDB, HDB, VDB, HARMobjectDB$}
\State{\texttt{Init $HARM$}}
\State{\texttt{Read $NDB$}}
\For{\texttt{$All~hosts$ in $NDB$}}
\State{\texttt{Add $host~to~HARM$}}
\State{\texttt{Read $HDB$}}
\For{\texttt{$All~vulnerability$ in $HDB$}}
\State{\texttt{Read $VDB$}}
\State{\texttt{$V \gets vulnerability$}}
\State{\texttt{Add $V~to~HARM$}}
\EndFor
\State{\texttt{ADD $Rechability~to~HARM$}}
\EndFor
\State{\texttt{Insert $HARM~to~HARMobjectDB$}}
\EndProcedure
\end{algorithmic}
\caption{Generating the HARM model}
\label{algorithm3}
\end{algorithm}

\begin{figure}
\center{\epsfig{file=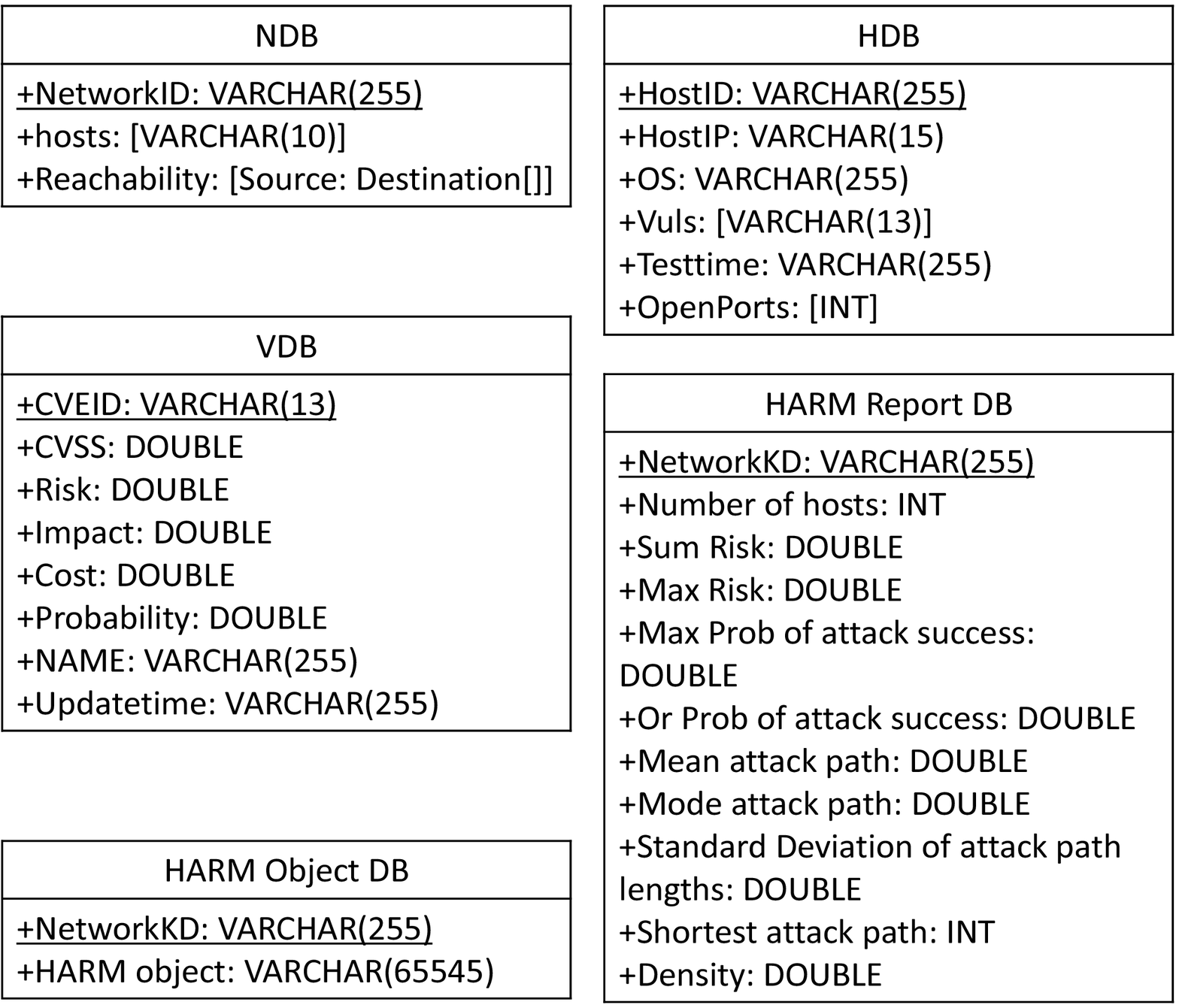, width=0.4\textwidth,clip=}}
\caption{DB schema}
\label{fig:DB} 
\end{figure}

Figure ~\ref{fig:DB} represents the entire DB schema. The framework used MongoDB and a NoSQL database. Because there are many multi-value elements in the framework, NoSQL database is easier to store than the RDB. However, the figure is shown using the RDB schema because there is no proper way to express the structure. NDB stores the VMs constituting the network and stores the reachability of each VM. HDB stores information about each VM. The main information is the IP address of the host, open ports, and the name of the vulnerability it has. The VDB stores the details of each vulnerability. CVSS, Risk, and Impact information is parsed by the NVD and stored in the database, and probability is obtained from the Common Vulnerability Scoring System used by the NVD. The attack cost value is saved as void because there is no way to calculate it automatically (i.e., it will rely on the monetary values of the assets in the cloud). Hence, the user can directly enter the cost value as appropriate.

\subsection{Phase 2}
In order to improve the security of the cloud, we can deploy security solutions based on the security assessment of the cloud we generated from Phase 1. However, it takes a great deal of resources to perform these tasks. Consequently, the service can also be stopped in the process. To solve this problem and test the security of the cloud, we propose Phase 2 that can create a new security model of the cloud by modifying the HARM configuration that was previously stored in the DB and evaluate their effectivenesses.

\begin{figure}
\center{\epsfig{file=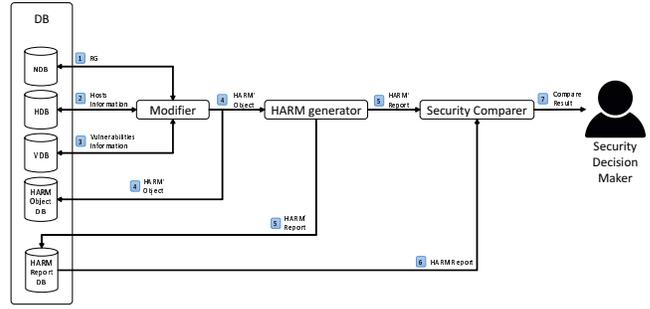, width=8.5cm,clip=}}
\caption{HARM modification and comparisons in Phase2}
\label{fig:Phase2} 
\end{figure}

Figure~\ref{fig:Phase2} shows the whole process of Phase 2. In this phase, NDB, HDB, and VDB are modified and saved in the DB. As a result, a new HARM model is generated, and the security analysis result of the new HARM model is provided to the user with a comparison report with the original HARM model generated in Phase 1. Phase 2 works recursively to change the cloud configurations in a direction that gradually improves the security posture of the cloud. To achieve this, different countermeasure solutions are compared and evaluated of their effectiveness by introducing security changes using the HARM. There are many different ways to deploy countermeasure selection to optimize the security \cite{chung2013nice}, but selecting the best countermeasure process is out of scope in this paper.

\begin{figure*}
\centering
\subfigure[Initial setup in Phase 1]
{
\epsfig{file=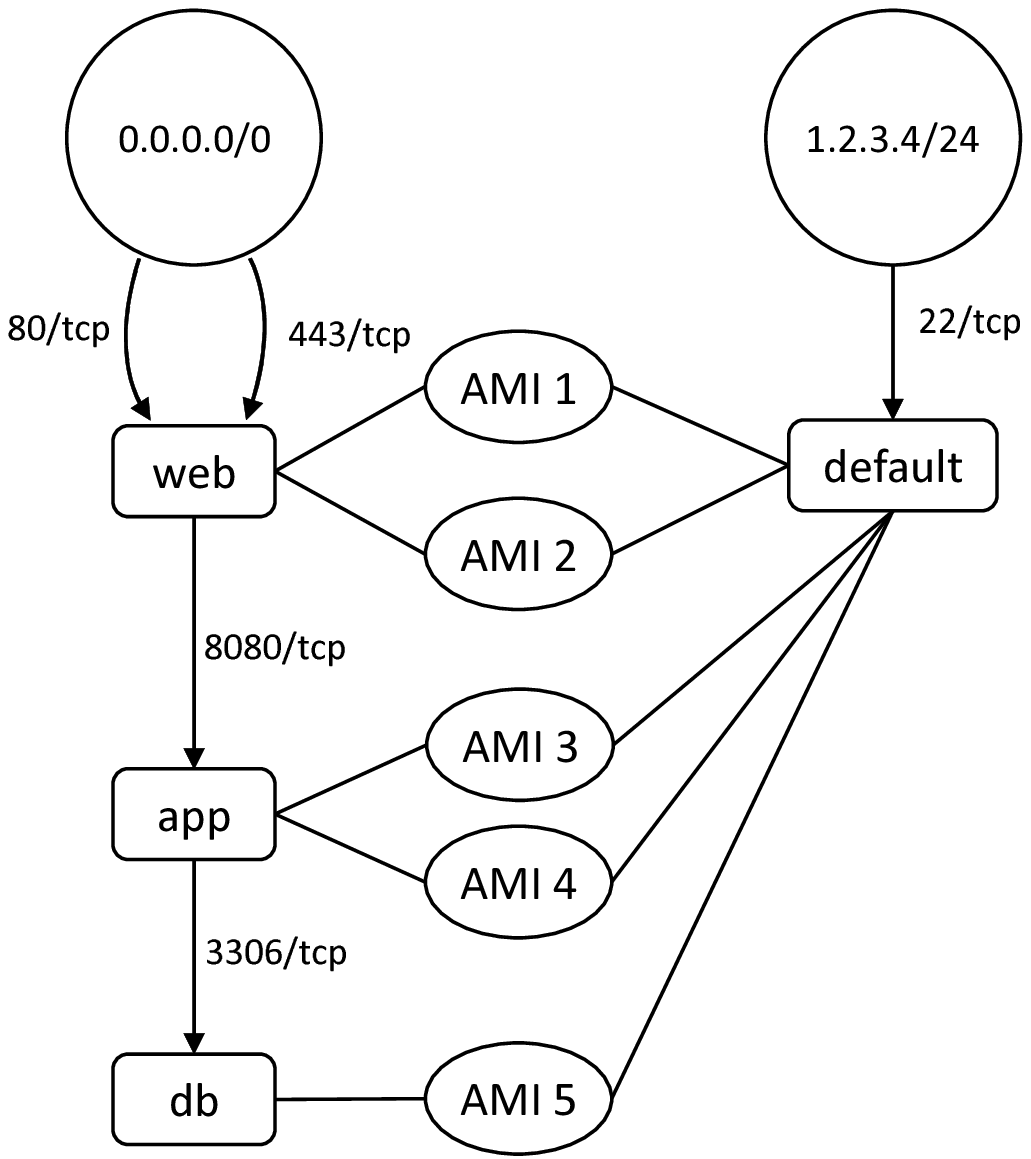, width=0.4\textwidth,clip=}
\label{SG(a)}
}
\subfigure[Modified setup after Phase 2]
{
\epsfig{file=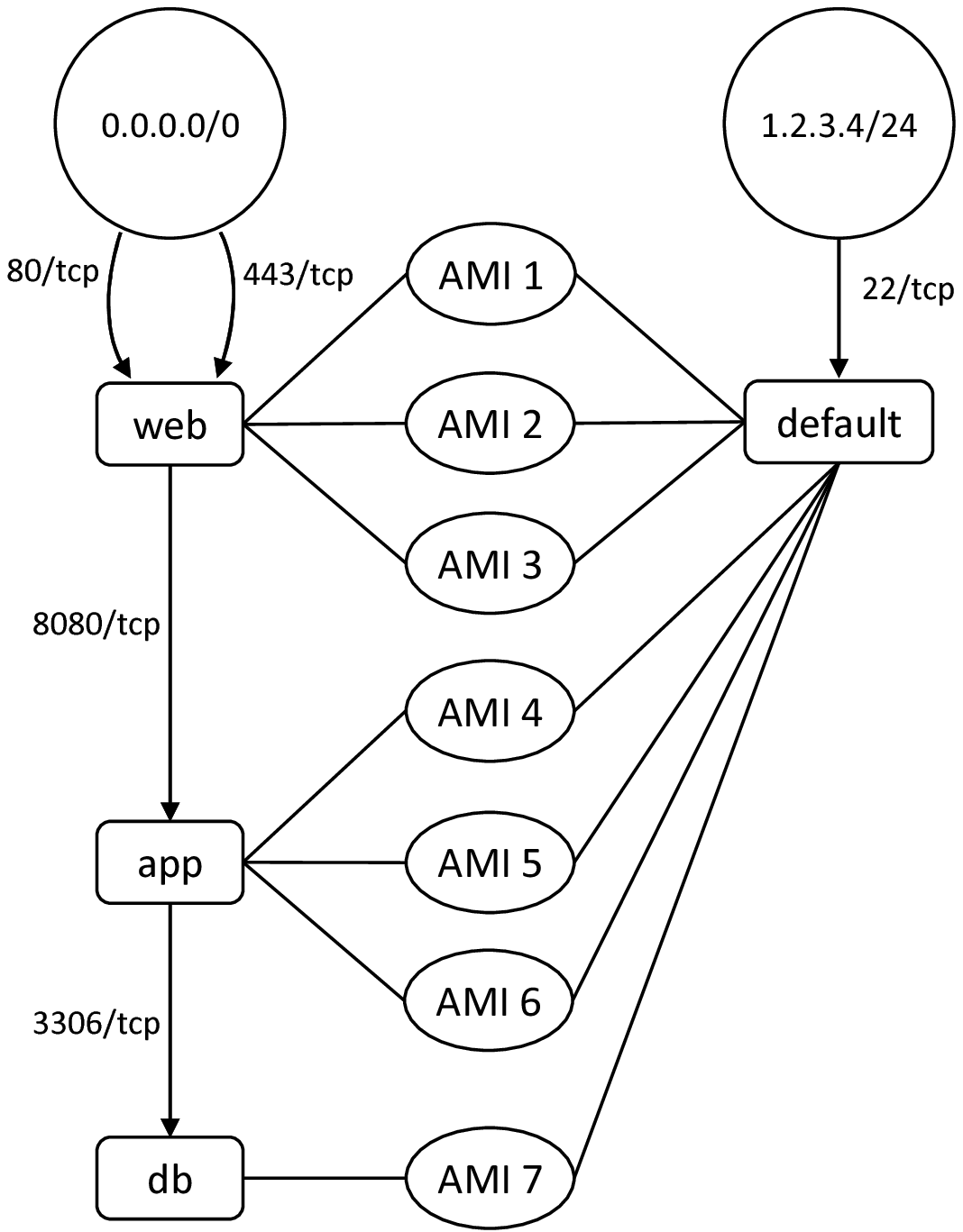, width=0.4\textwidth,clip=}
\label{SG(c)}
}
\caption{Relationship between AMIs and SGs in testbed 1}
\label{fig:SG1}
\end{figure*}

\section{Experimental Analysis}
\label{result}

For the experiment, we setup two different cloud testbeds to demonstrate the applicability and practical use of the CloudSafe tool. The two target clouds were implemented in the AWS. The first is implemented in three tiers of Web, App, and DB, and the second one is implemented as a streaming server. The configuration of the testbeds are destailed in the Appendix \ref{appendix}. We show how the two phases described in the previous section are applied in the two cloud testbeds in the following subsections.


\subsection{Applying Phase 1 in the testbeds}

Figure ~\ref{fig:SG1} shows the connections between the AMI and SG in testbed 1. The DB only allows connections from the App, and the App only uses connections from the Web. The Web allows connections via http and https. All AMIs allow connection of a specific IP from port 22, which is for using SSH.

\begin{figure}
\centering
\subfigure[Phase 1 generation]
{
\epsfig{file=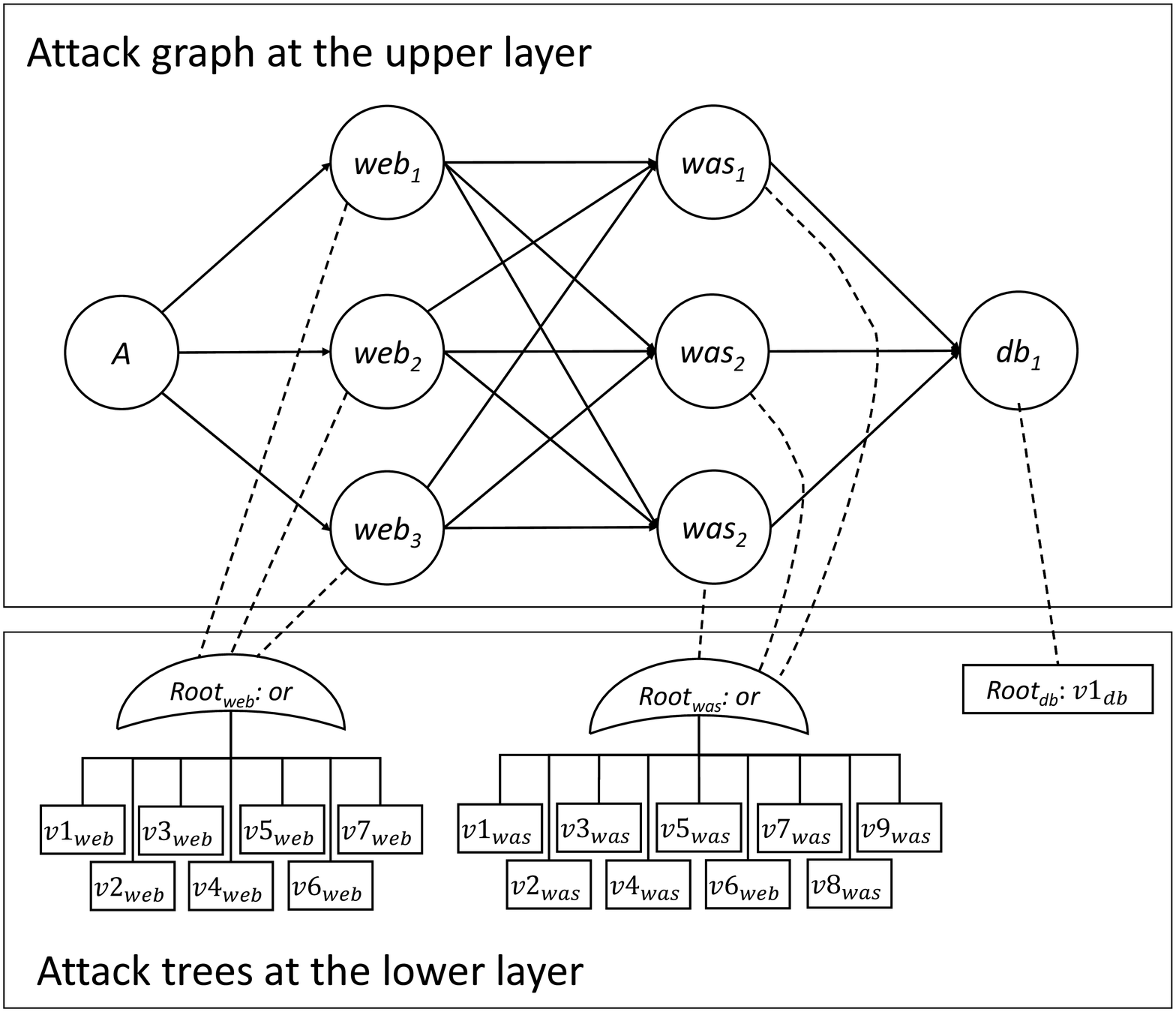, width=0.35\textwidth,clip=}
\label{s1HARM(a)}
}
\subfigure[Configuration changes made in Phase 2]
{
\epsfig{file=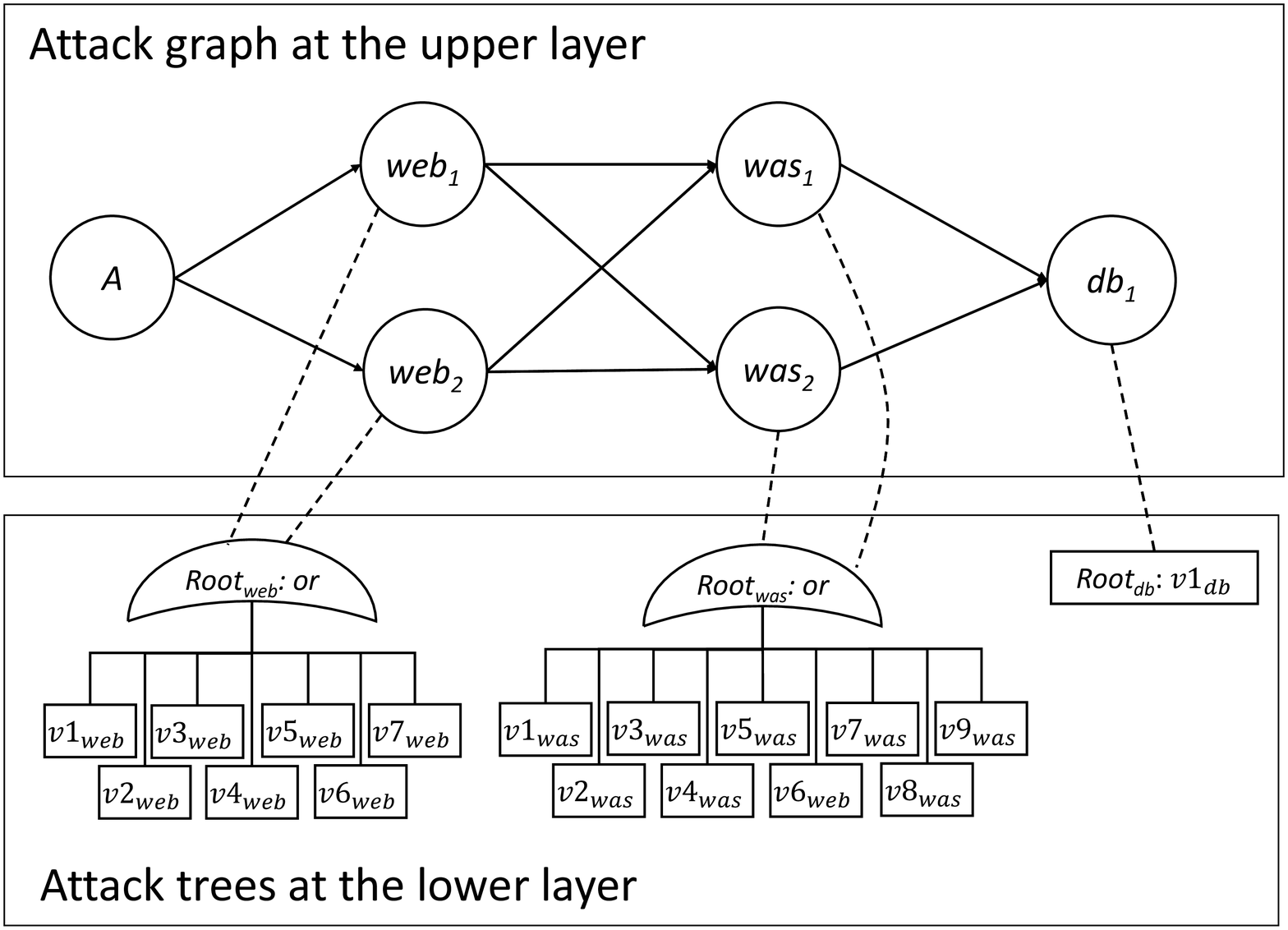, width=0.35\textwidth,clip=}
\label{s1HARM(c)}
}
\caption{HARM for testbed 1}
\label{fig:target1HARM}
\end{figure}


As an illustrative example, Figure~\ref{fig:target1HARM} shows the HARM model generated for testbed 1. Potential attack possible paths are calculated in HARM. Users can also retrieve graphical view of the HARM to understand the cloud components and their security relationships. We omit the illustrative example for testbed 2 due to the limited space.

\begin{table}
\caption{Security analysis results for testbed 1}
\label{HARM result1}
\begin{tabular}{lll}
\hline
\multicolumn{1}{c}{\multirow{2}{*}{Metrics}} & \multicolumn{2}{c}{Values} \\ \cline{2-3} 
 & Initial & Modified \\ \hline
Number of hosts & 5 & 7\\
Sum Risk & 617.346 & 274.376 \\
Max Risk & 146.223 & 64.988 \\
Or Probability of attack success & 0.965529 & 0.894082 \\
Max Probability of attack success & 0.934753 & 0.702739\\
Mean of attack path lengths & 3 & 3 \\
Mode of attack path lenghts & 3 & 3 \\
Standard Deviation of attack path lengths & 0 & 0 \\
Shortest attack path length & 3 & 3 \\
Density & 0.267857 & 0.266667 \\ \hline
\end{tabular}
\end{table}

\begin{table}
\caption{Security analysis results for testbed 2}
\label{HARM result2}
\begin{tabular}{lll}
\hline
\multicolumn{1}{c}{\multirow{2}{*}{Metrics}} & \multicolumn{2}{c}{Values} \\ \cline{2-3} 
 & Initial & Modified \\ \hline
Number of hosts & 5 & 7\\
Sum Risk & 276.701 & 86.33 \\
Max Risk & 73.435 & 24.694 \\
Or Probability of attack success & 0.80244 & 0.509267 \\
Max Probability of attack success & 0.93959 & 0.674334 \\
Mean of attack path lengths & 2.7 & 2.5 \\
Mode of attack path lenghts & 3 & 3 \\
Standard Deviation of attack path lengths & 0.16 & 0.25 \\
Shortest attack path length & 2 & 2 \\
Density & 0.238095 & 0.25 \\ \hline
\end{tabular}
\end{table}

\begin{figure}
\centering
\begin{subfigure}[]
{\epsfig{file=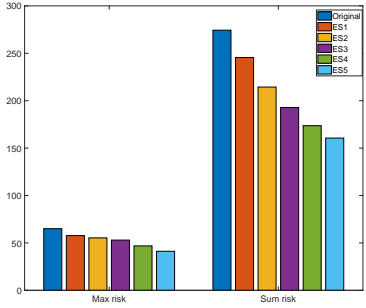, width=0.32\textwidth,clip=}\label{fig:S1r}}
\end{subfigure}
\begin{subfigure}[]
{\epsfig{file=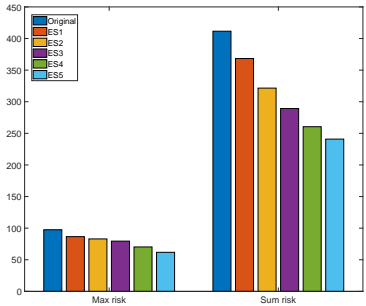, width=0.32\textwidth,clip=}\label{fig:S3r}}
\end{subfigure}
\begin{subfigure}[]
{\epsfig{file=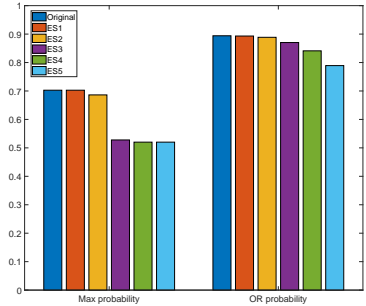, width=0.32\textwidth,clip=}\label{fig:S1p}}
\end{subfigure}
\begin{subfigure}[]
{\epsfig{file=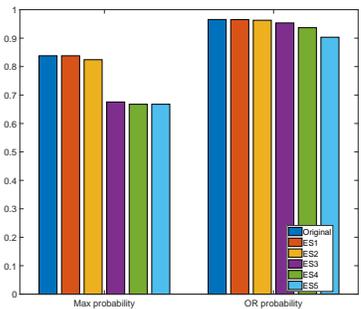, width=0.32\textwidth,clip=}\label{fig:S3p}}
\end{subfigure}
\caption{Changes in security posture when patching vulnerabilities using PSV. (a) and (b) show the risk report for testbed 1 and testbed 2, respectively. (c) and (d) show the probability of attack success report for testbed 1 and testbed 2, respectively.}\label{fig:psv}
\end{figure}

%
%
%
%
%
%

Table~\ref{HARM result1} shows the security analysis results for testbed 1, and Table \ref{HARM result2} for testbed 2. They include the security analysis results from both Phase 1 and Phase 2, which can be compared directly. The CloudSafe tool generates the security report using various security metrics, and other relevant metrics, not only security related (e.g., performance, reliability etc), can also be implemented as additional metric modules and integrated with the HARM. By generating those security reports, we can easily compare different security aspects of various cloud environments. One note is that using the HARM, the attack cost and the return on attack calculation functions require the attack cost to be specified in the DB. As NVD does not provide this cost, this function is not used in this study (however, it can be calculated if specified by the user). Next, we consider applying security countermeasures to compare how the security posture changes and evaluate the changes using CloudSafe.

\begin{table*}[ht!]
\caption{Phase1 execution time}
\centering
\label{time}
\begin{tabular}{lllll}
\hline
\multicolumn{1}{c}{\multirow{2}{*}{Experiment}} & \multicolumn{4}{c}{Values} \\ \cline{2-5} 
\multicolumn{1}{c}{} & \multicolumn{2}{l}{model1} & \multicolumn{2}{l}{model2} \\ \hline
Obtain SGs from Amazon & 1.72$s$ & 1.74$s$ & 1.69$s$ & 1.71$s$ \\ \hline
Parsing and build Reachability Graph & 0.48$s$ & 0.48$s$ & 0.46$s$ & 0.44$s$  \\ \hline
Insert and Update Database & 2.24$s$ & 2.22$s$  & 2.23$s$   &  2.28$s$  \\ \hline
Vulnerability Scanning(All Tcp) & 5$min$28$s$ &  &  &  \\ \hline
Scan report parsing & 1.35$s$ & 1.41$s$ & 1.28$s$ & 1.74$s$  \\ \hline
Insert and Update Host(AMI) Database  & 1.19$s$ & 1.15$s$ & 1.24$s$ & 1.12$s$ \\ \hline
Insert Vulnerability Database(Including NVD parsing) & 3.42$s$ & 3.69$s$ & 3.18$s$ & 3.89$s$ \\ \hline
\end{tabular}
\label{tab:harmresult1}
\end{table*}

\subsection{Applying Phase 2 in the testbeds}
An example countermeasure selection we deploy here is vulnerability patching, to show how the security posture changes using CloudSafe when countermeasures are deployed. One way to optimize the vulnerability patching is to prioritize the set of vulnerabilities to patch. To do this, we compute the prioritized set of vulnerabilities (PSV) as specified in \cite{hong2014what}. The PSV is a method that can be applied to improve the security of a network using the HARM. For our implementations, we use the exhaustive search (ES) algorithm method among other PSV algorithms. The ES algorithm uses the risk and cost values to prioritize vulnerabilities to be patched in the cloud testbeds. Table~\ref{Table:PSV} shows the PSV ranking of the testbeds, and the result of applying PSV to the testbeds is shown in Fig~\ref{fig:psv}. We eliminated the vulnerability sequentially from ES1 to ES5 respectively, and the results show the gradual reduction in the risk and the probability of attack success.

The result shows that for testbed 1 (i.e., Figures \ref{fig:S1r} and \ref{fig:S1p}), patching more vulnerabilities would decrease the risk and the probability of attack success in a consistent trend. That is, the amount of reduction is roughly equivalent when a vulnerability is patched everytime. On the other hand, the effectiveness of patching vulnerabilities is different with testbed 2, where the reductions in the risk and the probability of attack success is less effective compared with testbed 1. Although there is a sharp reduction when patching 3 vulnerabilities, there is no significant reduction when more vulnerabilities are patched. CloudSafe provides such security analysis for users to view and understand how the security posture can change when different countermeasures are applied in the cloud.

\subsection{Overhead}
To understand the performance overhead using CloudSafe, we measure the time take in Phase 1. The time taken using the CloudSafe tool on testbed 1 was measured and the mean value was used from 10 measurements. Table \ref{tab:harmresult1} shows the overhead of gathering the data to populate the DBs. It shows that most tasks can be completed reasonably quickly (within matter of few seconds), while the biggest bottleneck is the vulnerability scanning of VMs. One approach to improve the vulnerability scanning time is to reduce the port range to scan, but vulnerabilities outside the measurement range would not be included in the analysis. Another approach is to parallelize the vulnerability scanning using the cloud, which we will investigate in our future work. 

\begin{figure}
\centering
\center{\epsfig{file=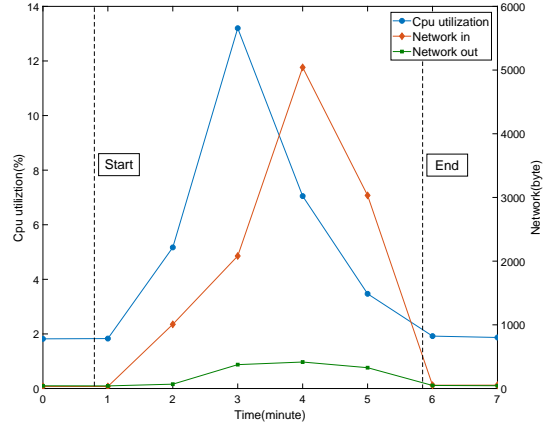, width=0.43\textwidth,clip=}}
\caption{VM resource overhead using CloudSafe}
\label{fig:test} 
\end{figure}

Next, we measure the computational overhead using CloudSafe. The use of the vulnerability analysis tools is the only factor that that affects the VM when using CloudSafe. Figure~\ref{fig:test} shows the load on the CPU and the network of the VMs when using CloudSafe. All TCP/IP ports were scanned and the load was measured in 1 minute increments. During the measurement, CPU usage increased by up to 11 percent, and the network had a maximum input of 5,042 bytes per minute and an output of 416 bytes. During the tool usage period, the cumulative input was 11,163 bytes and the output was 1,234 bytes. Network load was not large, but the CPU utilization increased significantly. There is need to adjust the scan options depending on the host's CPU performance or the role. On the other hand, disk reads and writes were almost unchanged.

\begin{table}
\caption{PSV using ES method for testbed 1 (top 6)}
\centering
\begin{tabular}{lllllll}
\hline
\multicolumn{1}{|l|}{Rank} & \multicolumn{1}{l|}{1} & \multicolumn{1}{l|}{2} & \multicolumn{1}{l|}{3} & \multicolumn{1}{l|}{4} & \multicolumn{1}{l|}{5} & \multicolumn{1}{l|}{6} \\ \hline
\multicolumn{1}{|l|}{ID} & \multicolumn{1}{l|}{v7web} & \multicolumn{1}{l|}{v2was} & \multicolumn{1}{l|}{v5web} & \multicolumn{1}{l|}{v6web} & \multicolumn{1}{l|}{v3web} & \multicolumn{1}{l|}{v7was} \\ \hline
\end{tabular}
\label{Table:PSV}
\end{table}

\section{Conclusion}
\label{conclusion}
Evaluating the security of the cloud can be a challenging task due to the scalability and dynamic nature, as well as the different privilege boundaries between the stakeholders (e.g., cloud service providers and clients). To address these problems, we proposed a framework to evaluate the security of the cloud using CloudSafe. CloudSafe provides semi-automated functions to collect and store the security information from the cloud, and also provide functions to modify the cloud configurations and compare how the security posture changes in the cloud. The results show that without too much user interventions, CloudSafe can collect security data from the cloud, evaluate the security posture of the cloud using the HARM, and generate reports that the user can utilize to assess the security of the cloud. Further, different countermeasures can be pre-evaluated prior to their deployment using CloudSafe to compare changes in the security posture of the cloud.

%

%

\section*{Acknowledgment}
This paper was made possible by Grant NPRP 8-531-1-111 from Qatar National Research Fund (QNRF). The statements made herein are solely the responsibility of the authors. 

\bibliography{main}

\section*{Appendix}
\label{appendix}
Table~\ref {Table1} shows the AWS resources and service versions used to setup the testbeds. Table ~\ref{Table2} shows the AWS resources used by the CloudSafe tool (WS refers to \textit{Windows Server}). The type of those instances were all t2.micro. Table ~\ref{Table:vuls} shows the vulnerability information of each AMI as a result of vulnerability scanning of the target cloud. 

\begin{table}[h]
\caption{AWS resources for the target clouds}
\centering
\begin{tabular}{llll}
\hline
\multicolumn{1}{c}{Instance name} & \multicolumn{1}{c}{OS} & \multicolumn{1}{c}{Spec}                                          & \multicolumn{1}{c}{Service} \\ \hline
Web server& \begin{tabular}[c]{@{}l@{}}Amazon\\ Linux\end{tabular} & \begin{tabular}[c]{@{}l@{}}vCPUs: 1\\ Memory: 1(GiB)\end{tabular} & Apache 2.4.1                \\ \hline
App server& \begin{tabular}[c]{@{}l@{}}Ubuntu\\ 14.04\end{tabular}& \begin{tabular}[c]{@{}l@{}}vCPUs: 1\\ Memory: 1(GiB)\end{tabular} & Tomcat 7.2.0                \\ \hline
DB server& \begin{tabular}[c]{@{}l@{}}Ubuntu\\ 14.04\end{tabular}  & \begin{tabular}[c]{@{}l@{}}vCPUs: 1\\ Memory: 1(GiB)\end{tabular} & MySQL 5.6.27                \\ \hline
FTP& \begin{tabular}[c]{@{}l@{}}Windows\\ 2013\end{tabular} & \begin{tabular}[c]{@{}l@{}}vCPUs: 1\\ Memory: 1(GiB)\end{tabular} & FileZilla 3.14.1                \\ \hline
Streamer& \begin{tabular}[c]{@{}l@{}}Windows\\ 2013\end{tabular}  & \begin{tabular}[c]{@{}l@{}}vCPUs: 1\\ Memory: 1(GiB)\end{tabular} & Wowza 4.3.0                \\ \hline
Vod bucket& \begin{tabular}[c]{@{}l@{}}Ubuntu\\ 14.04\end{tabular} & \begin{tabular}[c]{@{}l@{}}vCPUs: 1\\ Memory: 1(GiB)\end{tabular} & MySQL 5.6.27                \\ \hline
\label{Table1}
\end{tabular}
\end{table}

\begin{table}[h]
\caption{Security Assessment Cloud Resource}
\centering
\begin{tabular}{l|l|l}
\hline
\multicolumn{1}{c}{Instance name} & \multicolumn{1}{c}{OS} & \multicolumn{1}{c}{Spec} \\ \hline
\begin{tabular}[c]{@{}l@{}}DB generation\end{tabular} & \begin{tabular}[c]{@{}l@{}}WS 2016\end{tabular} & \begin{tabular}[c]{@{}l@{}}vCPUs: 1, Memory: 1(GiB)\end{tabular} \\ \hline
\begin{tabular}[c]{@{}l@{}}HARM \end{tabular} & \begin{tabular}[c]{@{}l@{}}WS 2016\end{tabular} & \begin{tabular}[c]{@{}l@{}}vCPUs: 1, Memory: 1(GiB)\end{tabular} \\ \hline
Database & Ubuntu 14.04 &\begin{tabular}[c]{@{}l@{}}vCPUs: 1, Memory: 1(GiB)\end{tabular} \\ \hline
\begin{tabular}[c]{@{}l@{}}Vuln. Scanner\end{tabular}  & Ubuntu 14.04 & \begin{tabular}[c]{@{}l@{}}vCPUs: 1, Memory: 1(GiB)\end{tabular} \\ \hline
\begin{tabular}[c]{@{}l@{}}Reachability Collector\end{tabular} & Ubuntu 14.04 & \begin{tabular}[c]{@{}l@{}}vCPUs: 1, Memory: 1(GiB)\end{tabular} \\ \hline
\end{tabular}
\label{Table2}
\end{table}

\begin{table}[h]
\caption{Vulnerabilities in AMIs}
\centering
\label{Table:vuls}
\begin{tabular}{|c|c|c|c|l}
\cline{1-4}
Vulnerability & CVE-ID & Probability & Risk &  \\ \cline{1-4}
$V1_{web}$ & CVE-2016-8740 & 0.5 & 1.45 &  \\ \cline{1-4}
$V2_{web}$ & CVE-2016-1546 & 0.43 & 1.849 &  \\ \cline{1-4}
$V3_{web}$ & CVE-2016-5387 & 0.51 & 3.264 &  \\ \cline{1-4}
$V4_{web}$ & CVE-2016-4979 & 0.5 & 1.45 & \\ \cline{1-4}
$V5_{web}$ & CVE-2016-6515 & 0.78 & 5.382 & \\ \cline{1-4}
$V6_{web}$ & CVE-2016-10009 & 0.75 & 4.8 & \\ \cline{1-4}
$V7_{web}$ & CVE-2015-8325 & 0.72 & 7.2 & \\ \cline{1-4}
$V1_{was}$ & CVE-2016-5388 & 0.51 & 3.264 & \\ \cline{1-4}
$V2_{was}$ & CVE-2016-3092 & 0.78 & 7.8 & \\ \cline{1-4}
$V3_{was}$ & CVE-2017-5647 & 0.5 & 1.45 & \\ \cline{1-4}
$V4_{was}$ & CVE-2017-5648 & 0.64 & 3.136 & \\ \cline{1-4}
$V5_{was}$ & CVE-2016-6816 & 0.68 & 4.352 & \\ \cline{1-4}
$V6_{was}$ & CVE-2016-8747 & 0.5 & 1.45 & \\ \cline{1-4}
$V7_{was}$ & CVE-2016-6515 & 0.78 & 6.9 & \\ \cline{1-4}
$V8_{was}$ & CVE-2016-10009 & 0.75 & 6.4 & \\ \cline{1-4}
$V9_{was}$ & CVE-2015-8325 & 0.72 & 7.2 & \\ \cline{1-4}
$V1_{db}$ & CVE-2013-2566 & 0.43 & 1.247 & \\ \cline{1-4}
$V1_{FTP}$ & CVE-2018-0087 & 0.56 & 1.247 & \\ \cline{1-4}
$V2_{FTP}$ & CVE-2018-5310 & 0.65 & 1.247 & \\ \cline{1-4}
$V3_{FTP}$ & CVE-2016-6515 & 0.78 & 5.382 & \\ \cline{1-4}
$V4_{FTP}$ & CVE-2016-10009 & 0.75 & 4.8 & \\ \cline{1-4}
$V5_{FTP}$ & CVE-2015-8325 & 0.72 & 7.2 & \\ \cline{1-4}
$V1_{streamer}$ & CVE-2018-7048 & 0.5 & 5.382 & \\ \cline{1-4}
$V2_{streamer}$ & CVE-2018-7049 & 0.43 & 5.382 & \\ \cline{1-4}
$V3_{streamer}$ & CVE-2018-16922 & 0.53 & 5.382 & \\ \cline{1-4}
$V4_{streamer}$ & CVE-2016-6515 & 0.78 & 5.382 & \\ \cline{1-4}
$V5_{streamer}$ & CVE-2016-10009 & 0.75 & 4.8 & \\ \cline{1-4}
$V6_{streamer}$ & CVE-2015-8325 & 0.72 & 7.2 & \\ \cline{1-4}
$V1_{bucket}$ & CVE-2013-2566 & 0.43 & 1.247 & \\ \cline{1-4}
\end{tabular}
\end{table}

\end{document}